\documentclass[a4paper]{revtex4}
\usepackage{graphicx}
\usepackage{fancyhdr}
\usepackage{amsmath}
\usepackage{color}

\begin{document}

\title{Collider Signatures of Gauge-Higgs Unification at LHC}

%

\author{Nobuhito Maru}
\affiliation{Department of Mathematics and Physics, Osaka City University, Osaka 558-8585, JAPAN}

\begin{abstract}
We have studied the Kaluza-Klein (KK) mode contributions to the diphoton decay of Higgs boson through the gluon fusion at LHC 
in a five dimensional $SU(3) \times U(1)'$ gauge-Higgs unification (GHU) model. 
The minimal matter content case is found that the diphoton decay is necessarily destructive comparing to the Standard Model (SM) prediction, 
which is unfortunately not supported by the current experimental data. 
Two extensions are considered to not only enhance the diphoton decay but also realize an observed Higgs mass.  
The $Z \gamma$ decay of Higgs boson is also studied 
and the striking result is found that there is no KK mode contributions to the decay at 1-loop level. 
This fact might be very useful to discriminate the GHU from other physics beyond the SM.  
\end{abstract}

\maketitle

\thispagestyle{fancy}


\section{Introduction}

A Higgs boson was discovered at LHC, but its origin is still a mystery. 
Gauge-Higgs unification (GHU) is one of the attractive scenarios beyond the Standard Model (SM). 
In this scenario, the SM Higgs boson is identified with an extra spatial component of the higher dimensional gauge field. 
As a result, the quantum correction to the Higgs mass is finite due to the gauge symmetry in higher dimensions 
 and the hierarchy problem is solved without supersymmetry. 
Furthermore, the interactions of the Higgs boson is controlled by the gauge principle 
 and there exists the Higgs couplings to new particles. 
We can therefore expect the deviations from the SM predictions and the collider signatures specific to the GHU.

\section{$gg \to H \to \gamma\gamma$}

We consider a five dimensional GHU model with $SU(3) \times U(1)'$ gauge groups, 
where the fifth dimension is compactified on an orbifold $S^1/Z_2$ with a compactification radius $R$ \cite{MO2}. 
By imposing the periodic boundary conditions for $S^1$ 
 and the $Z_2$ parity conditions with a matrix $P={\rm diag}(-,-,+)$ on the gauge fields, 
\begin{eqnarray}
A_\mu(-y) = P A_\mu(y) P^\dag, \qquad A_5(-y) =-P A_5(y) P^\dag
\end{eqnarray} 
where $\mu=0,1,2,3$ and $y$ denotes the fifth coordinate. 
These boundary conditions lead to the gauge symmetry breaking $SU(3) \to SU(2) \times U(1)$ 
 and the SM Higgs doublet is realized as the off diagonal elements of a zero mode of $A_5$. 

One of the embeddings of the SM fermions into $SU(3)$ representations is as follows. 
The $d,s,b$ ($u,c$) quarks are assigned to the fundamental (the two-rank symmetric tensor) 
representations, ${\bf 3}~({\bf 6}^*)$. 
The charged leptons $e, \mu, \tau$ are assigned to the three-rank totally symmetric tensor representations, ${\bf 10}$. 
Since Yukawa coupling is the gauge coupling in GHU, 
 the fermion masses lighter than the W-boson mass can be reproduced 
 by the exponentially suppressed overlap integrals of zero mode fermion wave functions 
 localized at different points in the fifth dimension. 
Only top quark is assigned to the four-rank totally symmetric tensor representation, ${\bf 15}^*$ 
 to obtain a top Yukawa coupling from the gauge coupling by the group theoretical enhancement factor. 

\begin{figure}[h]
\centering
\includegraphics[width=55mm]{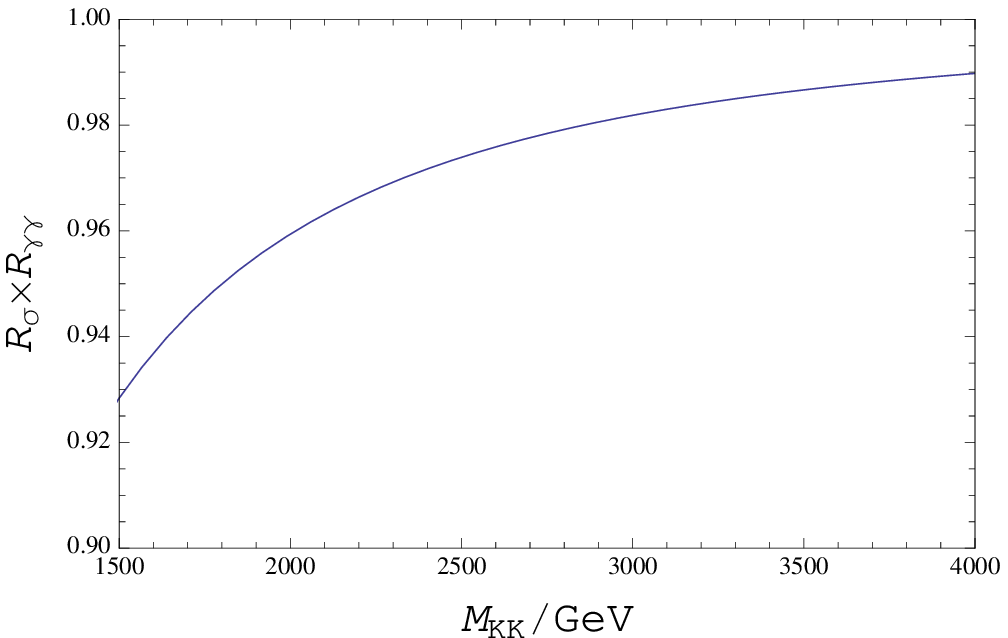}
\includegraphics[width=55mm]{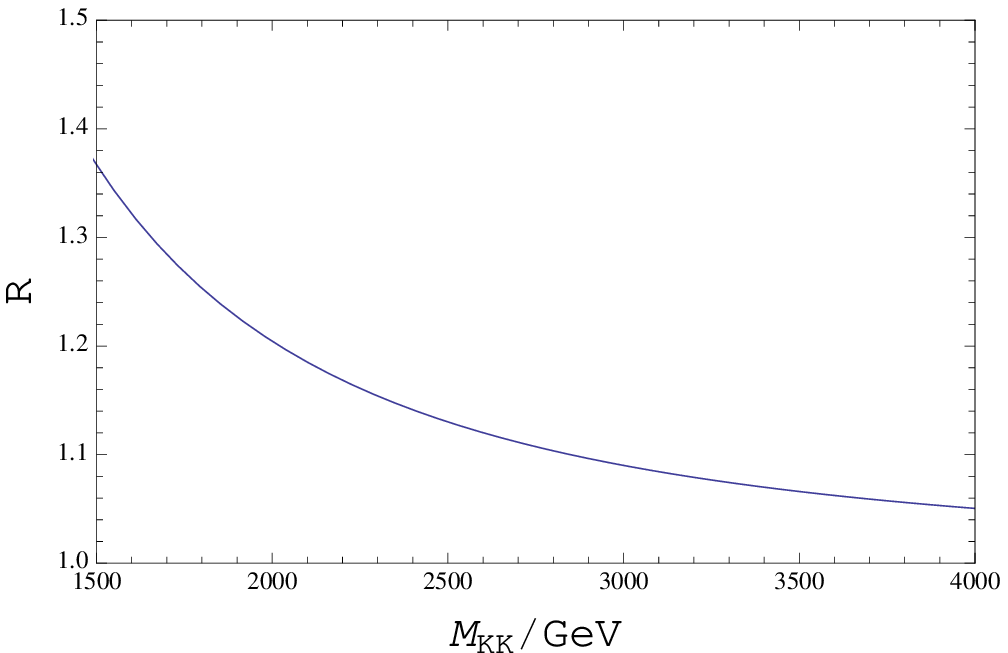}
\includegraphics[width=55mm]{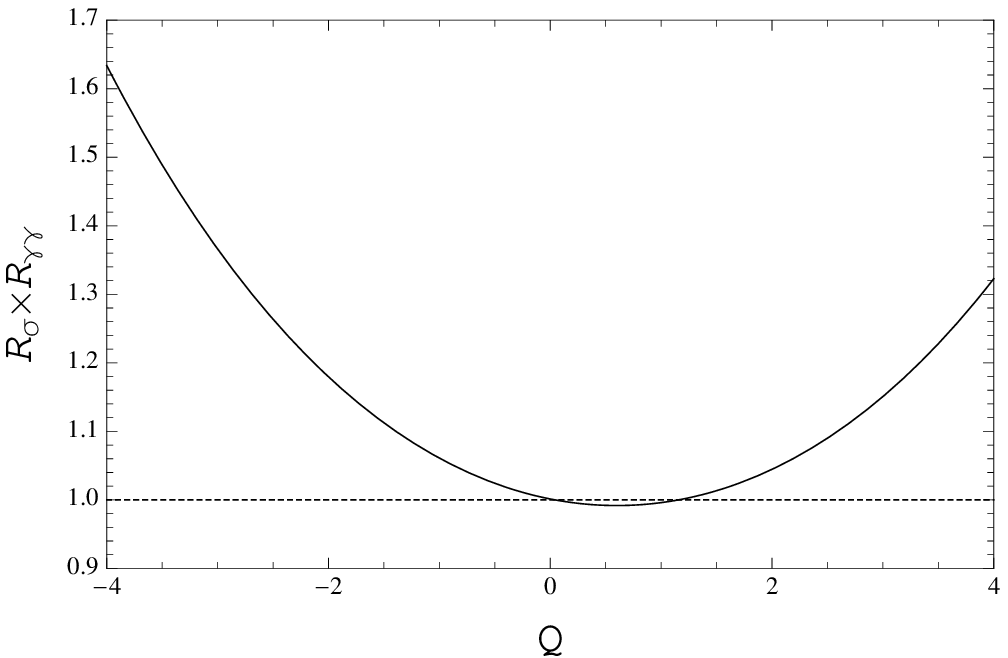}
\caption{The plots of the signal strength of $gg \to H \to \gamma\gamma$ in GHU without extra fermions (the left one) 
 and with a ${\bf 10}$ representation (the center and the right ones). 
The first two plots are written as a functions of the compactification scale $M_{KK}=1/R$, 
 but the last one as a function of the $U(1)'$ charge of ${\bf 10}$.} 
\label{plots}
\end{figure}

In this work, we focus on the gluon fusion $gg \to H$ and the diphoton decay mode $H \to \gamma\gamma$ at LHC. 
It is well known that the gluon fusion is dominantly generated by the top quark loop 
 and the diphoton decay is by the top and W-boson loops. 
These contributions are calculated from the coefficients of the following dimension five operators 
 ${\cal L}_{{\rm eff}} = h C_{gg} G^a_{\mu\nu} G^{a\mu\nu}, h C_{\gamma\gamma} F_{\mu\nu} F^{\mu\nu}$
where $h$ is the SM Higgs and $G_{\mu\nu}, F_{\mu\nu}$ are the gluon and photon field strength, respectively. 
The SM results are  
\begin{eqnarray}
C_{gg}^{{\rm SM top}} = \frac{\alpha_s}{12\pi v}, \qquad
C_{\gamma\gamma}^{{\rm SMtop}} = \frac{2\alpha_{em}}{9\pi v}, \qquad
C_{\gamma\gamma}^{{\rm SMW}} = -\frac{7\alpha_{em}}{8\pi v}. 
\label{SM}
 \end{eqnarray}
In GHU, the corresponding KK mode contributions should be taken into account as 
\begin{eqnarray}
&&C_{gg}^{{\rm KKtop}} = \frac{\alpha_s}{12\pi v} \sum_{n=1}^\infty 
 \left[ \frac{m_t}{m_n+m_t} -\frac{m_t}{m_n-m_t} \right] \simeq -\frac{\alpha_s}{32\pi v}(\pi m_t R)^2, \label{KKtop1} \\
&&C_{\gamma\gamma}^{{\rm KKtop}} = \frac{2\alpha_{em}}{9\pi v} 
\sum_{n=1}^\infty
\left[ \frac{m_t}{m_n+m_t} -\frac{m_t}{m_n-m_t} \right] \simeq -\frac{2\alpha_{em}}{27\pi v}(\pi m_t R)^2, \label{KKtop2} \\
&&C_{\gamma\gamma}^{{\rm KKW}} = 
 -\frac{7\alpha_{em}}{8\pi v} \sum_{n=1}^\infty 
 \left[ \frac{m_W}{m_n+m_W} -\frac{m_W}{m_n-m_W} \right] \simeq \frac{7\alpha_{em}}{24\pi v}(\pi m_W R)^2. 
\label{KKW}
\end{eqnarray}
Note that the KK mode contributions to the gluon fusion and the diphoton decay are finite 
 due to the gauge symmetry although each term is logarithmic divergent. 
In the final expressions of the above equations, 
 the approximations that the compactification scale is much larger than the top and W-boson masses are made. 
In fact, we can also check that the dimension five operators mentioned above are forbidden in GHU \cite{Maru}, 
 which supports the finiteness. 
The sign of the KK mode contributions are opposite to that of the SM predictions. 
From these results, the signal strength of $gg \to H \to \gamma\gamma$,  
\begin{eqnarray}
R \equiv R_\sigma \times R_{\gamma\gamma} = 
\left| \frac{C_{gg}^{{\rm SMtop}} + C_{gg}^{{\rm KKtop}}}{C_{gg}^{{\rm SMtop}}} \right|^2
\left| \frac{C_{\gamma\gamma}^{{\rm SMtop}} + C_{\gamma\gamma}^{{\rm SMW}} + C_{\gamma\gamma}^{{\rm KKtop}} 
+ C_{\gamma\gamma}^{{\rm KKW}}}{C_{\gamma\gamma}^{{\rm SMtop}} + C_{\gamma\gamma}^{{\rm SMW}}} \right|^2
\end{eqnarray}
is calculated. 
$R_\sigma (R_{\gamma\gamma})$ is the ratio of Higgs production cross section 
(the partial decay width of $H \to \gamma \gamma$) in our model to the SM one.  
The numerical result is shown in the left plot of FIG. \ref{plots}. 
We found that the minimal model of GHU predicts the signal strength of the diphoton decay of Higgs boson $R$ 
 is necessarily destructive comparing to the SM prediction, 
 which was a prediction before the operation of LHC \cite{MO1}. 
This is easily understood that the suppression via the gluon fusion process $gg \to H$ 
 is dominated over that of diphoton decay $H \to \gamma\gamma$. 
Unfortunately, the experimental data is unlikely to support this prediction. 

Therefore, we need an extension of our model to explain the data. 
A simple extension is to introduce an extra color singlet fermion.  
Such an extension is understood to be effective since the signal strength is more suppressed if the extra fermions have color charges 
 and the ratio of KK fermion contributions to $H \to \gamma \gamma$ 
 and the SM ones is necessarily positive. 
\begin{eqnarray}
C_{\gamma\gamma}^{{\rm KKfermions}} < 0, \quad 
C_{\gamma\gamma}^{{\rm SMtop}}+C_{\gamma\gamma}^{{\rm SMW}} < 0 \quad \to \quad  
\frac{C_{\gamma\gamma}^{{\rm KKfermions}}}{C_{\gamma\gamma}^{{\rm SMtop}}+C_{\gamma\gamma}^{{\rm SMW}}} > 0
\end{eqnarray}
 which can be seen from Eqs.~(\ref{SM}) and (\ref{KKtop2}).  
As a result, the signal strength of $gg \to H \to \gamma\gamma$ is enhanced. 

In \cite{MO2}, two examples of the extra fermion are considered, 
 namely ${\bf 10}$ and ${\bf 15}$ representations of $SU(3)$ with bulk masses and the half-periodic conditions. 
The reason of taking the half-periodic boundary condition is to avoid extra massless fermions absent in the SM. 
The mass of the lightest KK mode is indeed to be $1/(2R)$ even in the case of vanishing bulk mass.  
We have also shown in \cite{MO2} by using 1-loop renormalization group  equation for the Higgs quartic coupling 
 that these extra fermions are required to realize an observed 125 GeV Higgs mass 
 and the bulk mass is determined as a decoupling scale of the extra fermion contributions. 
 
The contributions of ${\bf 10}$ and {\bf 15} representations to $H \to \gamma\gamma$ are obtained as
\begin{eqnarray} 
&&C_{\gamma\gamma}^{{\rm KK10}} 
 = (Q-1)^2 F(3m_W) + (Q-1)^2 F(m_W) + Q^2 F(2m_W) + (Q+1)^2 F(m_W), \\ 
&&C_{\gamma\gamma}^{{\rm KK15}} = 
 (Q-4/3)^2 F(4m_W) + (Q-4/3)^2 F(2m_W) 
+ (Q-1/3)^2 F(3m_W) + (Q-1/3)^2 F(m_W) \nonumber \\
&& \hspace*{15mm} + (Q+2/3)^2 F(2m_W) + (Q+5/3)^2 F(m_W), \\
&&F(m_W) = \frac{\alpha_{em}}{6\pi v} m_W 
\sum_{n=0}^\infty 
\left( 
\frac{ m_{n+\frac{1}{2}}+m_W }{(m_{n+\frac{1}{2}}+m_W)^2+M^2} 
+
\frac{ m_{n+\frac{1}{2}}-m_W }{(m_{n+\frac{1}{2}}-m_W)^2+M^2} 
\right) 
\simeq
-\frac{\alpha_{em}}{6 \pi v} 
 \frac{(\pi m_W R)^2}{\cosh (\pi MR)} 
\end{eqnarray}
by noting the decompositions into $SU(2) \times U(1)$ representations 
 and the corresponding KK mass spectrum after the electroweak symmetry breaking
\begin{eqnarray}
 {\bf 10} = {\bf 4}_{1/2} \oplus {\bf 3}_{0} \oplus {\bf 2}_{-1/2} \oplus {\bf 1}_{-1} 
\end{eqnarray}
and
\begin{eqnarray}
&& 
\left( m_{n,-1}^{(\pm)} \right)^2 
 = \left( m_{n+\frac{1}{2}} \pm 3 m_W \right)^2 +M^2,~~ 
   \left( m_{n+\frac{1}{2}} \pm   m_W \right)^2 +M^2, \nonumber \\ 
&& 
\left( m_{n,0}^{(\pm)} \right)^2 
 = \left( m_{n+\frac{1}{2}} \pm 2 m_W \right)^2 +M^2,~~ 
   m_{n+\frac{1}{2}}^2 + M^2, \nonumber \\ 
&& 
\left( m_{n,+1}^{(\pm)} \right)^2 
 = \left( m_{n+\frac{1}{2}} \pm m_W \right)^2 +M^2,~~ 
\left( m_{n,+2}^{(\pm)} \right)^2 
 = m_{n+\frac{1}{2}}^2 +M^2
\end{eqnarray}  
for ${\bf 10}$ and
\begin{eqnarray}
  {\bf 15} = {\bf 5}_{2/3} 
 \oplus {\bf 4}_{1/6} \oplus {\bf 3}_{-1/3} \oplus {\bf 2}_{-5/6} \oplus {\bf 1}_{-4/3} 
\end{eqnarray}
and  
\begin{eqnarray}
&& \left( m_{n,-4/3}^{(\pm)} \right)^2 
 = \left( m_{n+\frac{1}{2}} \pm 4 m_W \right)^2 +M^2,~~ 
   \left( m_{n+\frac{1}{2}} \pm 2  m_W \right)^2 +M^2,~~  
   m_{n+\frac{1}{2}}^2+M^2, 
  \nonumber \\ 
&& 
\left( m_{n,-1/3}^{(\pm)} \right)^2 
 = \left( m_{n+\frac{1}{2}} \pm 3 m_W \right)^2 +M^2,~~~ 
   \left( m_{n+\frac{1}{2}} \pm   m_W \right)^2 +M^2, 
  \nonumber \\ 
&& 
\left( m_{n,2/3}^{(\pm)} \right)^2 
 = \left( m_{n+\frac{1}{2}} \pm 2 m_W \right)^2 +M^2,~~ 
   m_{n+\frac{1}{2}}^2+M^2,
  \nonumber \\ 
&&
\left( m_{n,5/3}^{(\pm)} \right)^2 
 = \left( m_{n+\frac{1}{2}} \pm m_W \right)^2 +M^2,~~
\left( m_{n,8/3}^{(\pm)} \right)^2 
 = m_{n+\frac{1}{2}}^2+M^2
 \end{eqnarray}
for ${\bf 15}$. 
$Q$ is $U(1)'$ charge of ${\bf 10}$ or ${\bf 15}$. 
$m_{n+\frac{1}{2}}=(n+\frac{1}{2})/R$.  
 
The results in the case of ${\bf 10}$ representation added are shown in the center and the right plots of FIG \ref{plots}.
The center plot represents the signal strength of $gg \to H \to \gamma\gamma$ as a function of the compactification scale. 
The $U(1)'$ charge of ${\bf 10}$ is chosen to be $ Q=-1$ as an example 
 and the bulk mass is fixed to be $M=0.55/R$ so as to realize 125 GeV Higgs mass. 
The right plot represents the signal strength of $gg \to H \to \gamma\gamma$ as a function of the $U(1)'$ charge with $1/R=3$ TeV fixed. 
It can be seen from both plots that the signal strength of $gg \to H \to \gamma\gamma$ is indeed enhanced 
 as we like by adjusting a free parameter $U(1)'$ charge. 
The $U(1)'$ charge of ${\bf 10}$ will be determined from the experimental data.  
The qualitatively similar results in the case of ${\bf 15}$ representation added is also obtained and see \cite{MO2}.  

A comment on a possibility of the dark matter for extra fermions is given. 
We note that the extra fermions ${\bf 10}$ and ${\bf 15}$ have a half-periodic boundary condition. 
Therefore, the lightest KK fermion is stable due to the boundary condition, which is regarded as some kind of $Z_2$ parity. 
If the experimental data indicate a vanishing $U(1)'$ charge of ${\bf 10}$ or ${\bf 15}$, 
 then the lightest KK fermion can be a good candidate of the dark matter. 


\section{$H \to Z \gamma$}

In the previous section, 
 we have seen the deviation of the diphoton decay of Higgs boson in GHU. 
This is caused by the contributions of the KK modes with electroweak charges. 
It is very natural that these KK mode contributions also affect the $Z \gamma$ decay of Higgs boson 
 and the deviation from the SM prediction is seen. 
From the phenomenological viewpoints, 
 the study of the model dependent correlations between the diphoton and $Z\gamma$ decays is very important 
 to discriminate the GHU model from the other physics beyond the SM. 
In \cite{MO3}, we have studied the $H \to Z\gamma$ process in GHU 
 and found a very striking result that we have no KK mode contributions to $Z\gamma$ decay at 1-loop in GHU. 
The reason is very simple. 
We cannot write any 1-loop diagram of $H \to Z\gamma$ since Higgs boson and the photon couple to the same mass eigenstate of KK fermions and KK W-bosons, 
 but the Z-boson does not. 
The relevant gauge interactions in the mass eigenstate are explicitly written as follows \cite{MO3}. 
\begin{eqnarray}
{\cal L} &\supset&  
\frac{g}{2} (\bar{\psi}_1^{(n)}, \bar{\tilde{\psi}}_2^{(n)}, 
\bar{\tilde{\psi}}_3^{(n)})  
\left(
\begin{array}{ccc}
\frac{2 \gamma_{\mu}}{\sqrt{3}} & W^{+}_{\mu} & W^{+}_{\mu} \\
W^{-}_{\mu} & - \frac{\gamma_{\mu}}{\sqrt{3}} & - Z_\mu  \\
W^{-}_{\mu} & -Z_\mu & - \frac{\gamma_{\mu}}{\sqrt{3}} 
\end{array}
\right) 
\gamma^{\mu} 
\left(
\begin{array}{c}
\psi_1^{(n)} \\
\tilde{\psi}_2^{(n)} \\
\tilde{\psi}_3^{(n)}
\end{array}
\right) \nonumber \\
&&+ 4 i Z_\mu \left( P_{\mu\nu}^{+(n)} N_\nu^{-(n)} - P_{\mu\nu}^{-(n)} N_\nu^{+(n)} + N_{\mu\nu}^{-(n)} P_\nu^{+(n)} - N_{\mu\nu}^{+(n)} P_\nu^{-(n)} \right) \nonumber \\
&&+ 8 \sqrt{3} i
\gamma_\nu Z_\mu 
\left(
- P_\mu^{-(n)} N_\nu^{+(n)} + N_\mu^{-(n)} P_\nu^{+(n)} - P_\mu^{+(n)} N_\nu^{-(n)} + N_\mu^{+(n)} P_\nu^{-(n)}
\right)
\end{eqnarray}
where the $\psi_1^{(n)}, \tilde{\psi}_2^{(n)}, \tilde{\psi}_3^{(n)}$ are the mass eigenstates 
 with mass eigenvalues $n/R, n/R+m, n/R-m$, respectively. 
$m$ is a fermion mass. 
$P^{\pm(n)}, N^{\pm(n)}$ are the mass eigenstates with mass eigenvalues $n/R+m_W, n/R-m_W$, respectively. 
Superscripts $\pm$ mean an electric charges. 
$P(N)_{\mu\nu} \equiv \partial_\mu P(N) \partial_\nu P(N) - \partial_\nu P(N) \partial_\mu P(N)$. 
Note that $\gamma_\mu$ in the last line is not the gamma matrices but at the photon. 
One can see that $Z$-boson couples to the different KK fermions and KK W-bosons, 
 which means that we cannot write any 1-loop diagram of $H \to Z\gamma$.  
This fact has never been seen as far as we know 
and might be very useful to discriminate the GHU from any other physics beyond the SM.

\section{Summary}

We have studied the KK mode contributions to the diphoton decay of Higgs boson at LHC, 
$gg \to H \to \gamma\gamma$, in a five dimensional $SU(3) \times U(1)'$ GHU model. 
These infinite number of contributions are finite due to the higher dimensional gauge symmetry and predictive. 
We have first found that the minimal GHU model yielding only the SM fermions as massless fermions predicts 
 a suppression comparing to the SM prediction and therefore cannot explain the experimental data \cite{MO1,MO2}. 
Two extensions of our model were considered \cite{MO2}. 
The extra color singlet fermions in the ${\bf 10}$ and ${\bf 15}$ representations of $SU(3)$ 
 with bulk masses and the half-periodic boundary condition are introduced. 
These extra fermions not only enhance the diphoton decay but also are required to realize the 125 GeV Higgs mass. 
$U(1)'$ charges of extra fermions are determined by the experimental data 
 and the stable lightest KK fermion can be a candidate of the dark matter 
 if the vanishing $U(1)'$ charge for ${\bf 10}$ or ${\bf 15}$  is supported by the data. 

The $Z \gamma$ decay of Higgs boson is also studied 
and the striking result is found that there is no KK mode contributions to the decay at 1-loop level \cite{MO3}. 
The reason is that we cannot write any 1-loop diagram of $H \to Z\gamma$ 
 since Higgs boson and the photon couple to the same KK mass eigenstates, but the $Z$-boson does not. 
This fact has never been seen as far as we know 
and might be very useful to discriminate the GHU from other physics beyond the SM.  

\begin{acknowledgments}
This work is supported in part by the Grant-in-Aid 
 for Scientific Research from the Ministry of Education, 
 Science and Culture, Japan No. 24540283.
\end{acknowledgments}

\bigskip 

\end{document}